\documentclass[showpacs,twocolumn,prl]{revtex4}
\usepackage{graphicx, amsmath, amsthm, amssymb}
\newcommand{\be}{\begin{equation}}
\newcommand{\ee}{\end{equation}}
\newcommand{\bea}{\begin{eqnarray}}
\newcommand{\eea}{\end{eqnarray}}
\newcommand{\bwt}{\begin{widetext}}
\newcommand{\ewt}{\end{widetext}}
\newcommand{\ra}{\rangle}
\newcommand{\la}{\langle}
\newcommand{\up}{\uparrow}
\newcommand{\dn}{\downarrow}
\newcommand{\vphi}{\varphi}
\newcommand{\dg}{\dagger}
\newcommand{\bk}{{\bf k}}
\newcommand{\om}{\omega}
\newcommand{\G}{{\cal G}}
\newcommand{\cD}{{\cal D}}

\newcommand{\sy}{{\hat\sigma}_{y}}

\newcommand{\uk}{{\it u}_{\bf k}}
\newcommand{\vk}{{\it v}_{\bf k}}
\newcommand{\sro}{Sr$_{2}$RuO$_{4}$ }
\begin{document}
\title{Hall Conductivity in a Spin-Triplet Superconductor}
\author{Wonkee Kim$^{1,2}$, F. Marsiglio$^{2}$, and C. S. Ting$^{1}$}
\affiliation{
$^1$ Texas Center for Superconductivity, Houston, Texas 77204
\\
$^2$Department of Physics, University of Alberta,
Edmonton, Alberta, Canada, T6G~2J1
}

\begin{abstract}
We calculate the Hall conductivity for a spin-triplet superconductor,
using a generalized pairing symmetry dependent on an arbitrary phase,
$\vphi$. A promising candidate for such an order parameter is Sr$_{2}$RuO$_{4}$,
whose superconducting order parameter symmetry is still
subject to investigation.
The value of this phase can be determined
through Kerr rotation and DC Hall conductivity measurements. Our calculations
impose significant constraints on $\vphi$.
\end{abstract}


\date{\today}
\maketitle

Superconducting strontium ruthenate (Sr$_{2}$RuO$_{4}$) \cite{maeno}
is remarkable for a variety of reasons: it is a layered compound without copper, and,
while its transition temperature is relatively low, the symmetry of the superconducting
order parameter is most certainly non-conventional \cite{mackenzie}.
It is believed\cite{ishida} that \sro is a spin-triplet superconductor\cite{sigrist};
thus, the orbital part of the Cooper pair should have the odd parity\cite{nelson}.
Moreover,
muon spin-relaxation measurements indicate that the superconducting state
in \sro breaks time-reversal symmetry\cite{luke}.
The diffraction patterns of Josephson junctions made from \sro
also illustrate this phenomenon\cite{kidwingira}.
In the interpretation of these experimental observations,
a spin-triplet superconductor with a $(k_{x}\pm ik_{y})$-wave gap symmetry
has been used\cite{mackenzie}.
On the other hand, characteristics of
the penetration depth\cite{bonalde} in \sro are not consistent with a
pure nodeless $p$-wave gap. Furthermore,
ultrasound attenuation measurements\cite{lupien}
in the superconducting state of \sro exclude the possibility
of a nodeless $p$-wave gap, while they seem to imply
a possible fourfold gap modulation. In this sense,
an $f$-wave gap has been also proposed\cite{graf}.

Recently, Xia {\it et al.}\cite{xia} have observed a Kerr rotation
developing in \sro as the temperature is lowered below $T_{c}=1.5K$.
They tried to understand this observation based on a theoretical
analysis\cite{yip} using a nodeless $p$-wave gap. However, their
theoretical estimate gives a Kerr angle of the order of
$10^{-3}$ nanorad, while the measured value is as big as $65$ nanorad.
Since the Kerr angle is proportional to the imaginary part of
the Hall conductivity\cite{xia}, Yakovenko\cite{yakovenko}
derived a Chern-Simons like term in the action associated
with the Hall conductivity, and estimated a Kerr angle of about $230$ nanorad.
In Ref\cite{yakovenko}, however,
the supercurrent (or Cooper pair) contribution to the Hall conductivity
is ignored. Another difficulty with this theory is
that the Kerr angle obtained by
Yakovenko is proportional to the square of the energy gap while
Xia {\it et al.} have observed that it is linear with the gap.

As explained in Ref.\cite{sigrist}, the superconducting state is described by
a linear combination of the basis functions for a given representation.
For a system with tetragonal symmetry such as \sro the superconducting gap can be
written in terms of the two-dimensional representation; thus
its momentum dependence would be $\eta_{x}k_{x}+\eta_{y}k_{y}$, where $\eta_{x,y}$
are complex numbers. Introducing a relative phase $\vphi$ between $\eta_{x}$ and
$\eta_{y}$, one can
express ${\vec\eta}$ as $(1,\;e^{i\vphi})$. Based on the Ginzburg-Landau (GL)
theory, we examine $\eta_{x,y}$ further. The corresponding fourth-order terms\cite{sigrist}
in the GL free energy would be $\beta_{1}({\vec\eta}\cdot{\vec\eta}^{*})^{2}+
\beta_{2}({\vec\eta}\times{\vec\eta}^{*})^{2}+\beta_{3}|\eta_{x}|^{2}|\eta_{y}|^{2}$ with $\beta_{1}>0$.
Depending on $\beta_{2}$ and $\beta_{3}$, we see what values of $\vphi$ would be possible.
For example, if $\beta_{2}>0$ and $4\beta_{2}>\beta_{3}$, $\vphi=\pm\pi/2$. When, however,
$\beta_{2}=0$ and $\beta_{3}<0$, one can consider an arbitrary value of
the relative phase $\vphi$.

In this letter we propose a generalized $p$-wave (we also consider $f$-wave)
gap with a relative phase $(\vphi)$ between the momenta along the $x$ and $y$
directions; namely
$\Delta_{\bk}=\Delta_{0}({\hat k}_{x}+e^{i\vphi}{\hat k}_{y})$ for the $p$-wave gap
and $\Delta_{\bk}=\Delta_{0}({\hat k}_{x}+e^{i\vphi}{\hat k}_{y})
({\hat k}^{2}_{x}-{\hat k}^{2}_{y})$ for the $f$-wave gap. We
derive an expression for the Hall conductivity and show that the Kerr angle is indeed proportional to
$\Delta_{0}$ as experimentally observed. As in the phenomenological model used by
Xia {\it et al.}, our derivation reveals that impurity scattering plays an important
role in the problem. The actual value of $\vphi$ we use can be identified by comparison
with experimental results for the Kerr angle.
The DC Hall conductivity at zero temperature is also computed because it is less sensitive
to impurity scattering but demonstrates a strong $\vphi$ dependence. Consequently,
the DC Hall conductivity would be another ideal experiment to determine $\vphi$.
We also discuss the chirality\cite{volovik,furusaki}
and the density of states (DOS) for these gaps.
The DOS of the $p$-wave gap provides a mapping of $\vphi$ onto a tiny gap\cite{miyake} associated
with the shape of the Fermi surface in Sr$_{2}$RuO$_{4}$.

We start with the current operator ${\bf j}$
\be
{\bf j}
= \frac{1}{2}e\sum_{k}{\bf v}_{k}{\hat\psi}^{\dg}_{k}{\hat\psi}_{k}
\ee
where
${\hat\psi}^{\dg}_{k} = \left(C^{\dg}_{k\up}\; C^{\dg}_{k\dn}\;
C_{-k\up}\; C_{-k\dn}\right)$.
Following the standard formalism we obtain
the current-current correlation ${\bf \Pi}(i\Omega)$ in the Matsubara representation
as follows:
\be
{\bf \Pi}(i\Omega)=\frac{1}{4}e^{2}\sum_{k}{\bf v}_{k}{\bf v}_{k}
T\sum_{\om}\mbox{Tr}\left[\G_{\bk}(i\om+i\Omega)\G_{\bk}(i\om)\right]
\ee
For a spin-triplet superconductor\cite{sigrist},
it is necessary to introduce the
$(4\times4)$ Green function $\G(\bk,i\om)$
\be
\G= \left ( \begin{array}{cc}
{\hat G} & -{\hat F} \\
-{\hat F}^{\dg} & -{\hat G}^{\dg}\end{array} \right )
\ee
with
\bea
{\hat G}(\bk,i\om) &=& -\frac{i\om+\xi_{\bk}}{\om^{2}+E^{2}_{\bk}}{\hat 1}
\nonumber\\
{\hat F}(\bk,i\om) &=& \frac{{\hat\Delta}_{\bk}}{\om^{2}+E^{2}_{\bk}}
\nonumber
\eea
where $\xi_{\bk} = \bk^{2}/2m-\epsilon_{F}$,
${\hat\Delta}_{\bk}=i({\bf d}(\bk)\cdot{\hat{\bf\sigma}})\sy$,
${\hat 1}$ is the $(2\times2)$ unit matrix,
and $E_{\bk}=\sqrt{\xi^{2}_{\bk}+{\mbox {tr}}[{\hat\Delta}_{\bk}
{\hat\Delta}^{\dg}_{\bk}]/2}$.
Since ${\hat\Delta}_{\bk}{\hat\Delta}^{\dg}_{\bk}=|{\bf d}(\bk)|^{2}{\hat 1}
+i[{\bf d}(\bk)\times{\bf d}^{*}(\bk)]\cdot{\hat{\bf\sigma}}$,
depending on ${\bf d}(\bk)\times{\bf d}^{*}(\bk)$, the pairing state is called
unitary if ${\bf d}(\bk)\times{\bf d}^{*}(\bk)=0$; otherwise it is
non-unitary. It is commonly assumed\cite{mackenzie} that
the unitary state is relevant to
\sro and ${\bf d}(\bk)=\Delta_{\bk}{\hat z}$:
\be
{\hat\Delta}_{\bk} = \left ( \begin{array}{cc}
0 & \Delta_{\bk} \\
\Delta_{\bk} & 0 \end{array} \right )
\ee
For this state,
the net spin average of a Cooper pair
${\mbox {tr}}[{\hat\Delta}^{\dg}_{\bk}{\hat{\bf\sigma}}{\hat\Delta}_{\bk}]=0$,
and $E_{\bk}=\sqrt{\xi^{2}_{\bk}+|\Delta_{\bk}|^{2}}$.
It is also possible to represent the $d$-wave gap
in the $(4\times4)$ matrix formalism as follows: ${\hat\Delta}_{\bk} = i\Delta_{\bk}\sy$
with $\Delta_{\bk} = \Delta_{0}({\hat k}^{2}_{x}-{\hat k}^{2}_{y})$.

Defining
$G(\bk,i\om)={\hat G}_{11}$ and $F(\bk,i\om)={\hat F}_{12}$,
the $xy$ component of the current-current correlation becomes,
at the bare bubble level,
\bwt
\be
\Pi_{xy}(i\Omega) = e^{2}\sum_{k}v_{x}v_{y}
T\sum_{\om}\left[G(\bk,i\om+i\Omega)G(\bk,i\om)+
F(\bk,i\om+i\Omega)F^{*}(\bk,i\om)\right].
\ee
\ewt
Using the symmetry of $\Pi_{xy}(i\Omega)$, one can see $\Pi_{xy}(i\Omega)=0$
for a pure nodeless $p$-wave gap.
A similar analysis has been done for order parameters with various symmetries\cite{li}.
The Hall conductivity follows readily from this expression:
$\sigma_{xy}(\Omega)=\frac{i}{\Omega}\Pi_{xy,ret}(\Omega)\equiv
\sigma'_{xy}(\Omega)+i\sigma''_{xy}(\Omega)$.
Introducing the spectral functions
${\cal A}(\bk,\om) = -2\mbox{Im}[G_{ret}(\bk,i\om)]$
and ${\cal B}(\bk,\om) = -2\mbox{Im}[F_{ret}(\bk,i\om)]$,
one obtains
\bea
\sigma''_{xy}(\Omega) &=&\frac{e^{2}}{\Omega}\sum_{k}v_{x}v_{y}\int{}
\frac{d\om' d\om''}{(2\pi)^{2}}\frac{f(\om'')-f(\om')}
{\om''-\om'+\Omega}
\nonumber\\
&\times&\left[{\cal A}(\bk,\om'){\cal A}(\bk,\om'')+
{\cal B}(\bk,\om'){\cal B}^{*}(\bk,\om'')\right]\;.
\label{freqeq}
\eea

In the clean limit, the spectral functions are
\bea
{\cal A}(\bk,\om)&=&2\pi|\uk|^{2}\delta(\om-E_{\bk})+
2\pi|\vk|^{2}\delta(\om+E_{\bk})
\nonumber\\
{\cal B}(\bk,\om)&=&2\pi\uk\vk\left[
\delta(\om+E_{\bk})-\delta(\om-E_{\bk})\right]\;,
\eea
where $\uk = \sqrt{(1+\xi_{\bk}/E_{\bk})/2}$ and
$\vk = \sqrt{(1-\xi_{\bk}/E_{\bk})/2}$. In this instance,
$\sigma''_{xy}(\Omega)$ vanishes regardless of the gap symmetry
because the ${\cal A}{\cal A}$ term (quasiparticle contribution)
is exactly canceled by the ${\cal B}{\cal B}$ term
(Cooper pair contribution). Nonetheless,
impurity scattering prevents a complete cancellation.
With impurity scattering rate $\gamma$, the corresponding
spectral functions can be approximated by \cite{schrieffer}
\bea
{\cal A}(\bk,\om)&=&|\uk|^{2}\cD(\om-E_{\bk})+
|\vk|^{2}\cD(\om+E_{\bk})
\nonumber\\
{\cal B}(\bk,\om)&=&\uk\vk\left[
\cD(\om+E_{\bk})-\cD(\om-E_{\bk})\right]\;,
\eea
where
\be
\cD(\om\pm E_{\bk})=\frac{2\gamma}{(\om\pm E_{\bk})^2+\gamma^{2}}.
\ee
When the self-energy $\Sigma$ due to impurity scattering is considered,
$\om\rightarrow{\tilde\om}=\om-\Sigma$.
Using the Born approximation for simplicity, one can readily evaluate the
frequency integrals in Eq. (\ref{freqeq}). Taking the low temperature and high
frequency limits, $T\rightarrow0$ (or $T\ll\Delta_{0}$),
and $\Omega\gg\Delta_{0}\gg\gamma$, as in Ref.\cite{xia}, we obtain
\be
\sigma''_{xy}(\Omega)\simeq\frac{e^{2}}{2\pi}\frac{\gamma}{\Omega^{3}}
\sum_{k}v_{x}v_{y}\ln\left[
1+\frac{\Omega^{4}-2\Omega^{2}(E^{2}_{\bk}-\gamma^{2})}
{(E^{2}_{\bk}+\gamma^{2})^{2}}\right].
\label{phinum}
\ee
Changing the summation to an integration over ${\bk}$, we arrive at the
high frequency result,
\be
\sigma''_{xy}(\Omega)\simeq\frac{e^{2}}{2\pi}v^{2}_{f}N(0)
\frac{\gamma\Delta_{0}}{\Omega^{3}}\;I(\vphi),
\ee
where $v_{f}$ is the Fermi velocity, $N(0)$ the DOS of the normal state, and
$I(\vphi)$ is, for the $p$-wave gap,
\begin{center}
$
I(\vphi)=\frac{4\cos(\vphi)}{3-3|\sin(\vphi)|}
\sqrt{\frac{2}{1+|\sin(\vphi)|}}\left[E(\nu)-|\sin(\vphi)|K(\nu)\right]
$
\end{center}
where $\nu = (1-|\sin(\vphi)|)/(1+|\sin(\vphi)|)$,
$K$ and $E$ are the complete elliptic integrals of the first and the second kind.
The corresponding result for the $f$-wave gap is
\begin{center}
$
I(\vphi)= \frac{8\sqrt{2}}{15}
\frac{1 + \frac{3}{2} |\sin(\vphi)|} {1 + |\sin(\vphi)|}
\Bigl[|\sin(\vphi/2)| - |\cos(\vphi/2)|\Bigr]
$.
\end{center}

In Fig.~1, we plot $I(\vphi)$ for the $p$-wave and $f$-wave gap.
Using a numerical integration of Eq. (\ref{phinum}) with actual values of $\Omega$ and $\gamma$
taken from experiment \cite{mackenzie,xia} gives results indistinguishable from those
in Fig.~1.
Note that $I(\vphi)$ changes its sign depending on $\vphi$ and vanishes
at $\vphi=\pi/2$ and $3\pi/2$. From this plot, we estimate
$\vphi\approx \pi/\sqrt{2}\sim 5\pi/6$, for which $I(\vphi)\sim 1$. These estimates
will depend on the precise symmetry as clearly the $p$-wave gap gives greater values
than the $f$-wave gap.
Since $v^{2}_{f}N(0)\sim\Omega$ from parameters in
Ref.\cite{mackenzie}, we obtain
\be
\sigma''_{xy}(\Omega)\approx
\frac{e^{2}}{2\pi}\frac{\gamma\Delta_{0}}{\Omega^{2}}
\label{imag_Hall_cond}
\ee
Note that our result is reduced from that in Ref.\cite{yakovenko} by
a factor of $\gamma/\Delta_{0}$.
This revises the theoretical estimate for the Kerr angle to $10 \sim 80$ nanorad
for $\gamma/\Delta_{0}\approx0.05 \sim 0.4$.
Its value congruous to the measured Kerr angle would be $0.15 \sim 0.35$.
Consequently, Eq.~(\ref{imag_Hall_cond}) correctly illustrates
the observed linear dependence of the Kerr angle on the gap value, and
supports the phenomenological expression used in Ref.\cite{xia}.

Because of the important role of impurity scattering,
the Kerr angle measurement may not be the ideal quantity
to determine the phase $\vphi$. Moreover, the dependence
of $\sigma''_{xy}(\Omega)$ on $\vphi$ is not sufficiently decisive
to pinpoint $\vphi$ accurately.
Since the low-lying quasiparticles behave sensitively
to $\vphi$, the DC Hall conductivity at zero temperature
can be a good measurement to determine the phase.
In the DC, low temperature limit we get for the Hall conductivity,
\be
\sigma'_{xy}(0) = \sigma_{0}\left\la
\frac{\sin(2\theta)}
{\left[\Delta^{2}_{\theta}+(\gamma/\Delta_{0})^{2}\right]^{3/2}}
\right\ra_{FS}\;,
\ee
where $\sigma_{0} = e^{2}v_{f}N(0)\gamma^{2}/(\pi\Delta^{3}_{0})$,
$\Delta_{\theta}=\Delta_{\bf k}/\Delta_{0}$, and
$\la\cdots\ra$ means the average over the Fermi surface.
Fig.~2 shows the normalized DC Hall conductivity,
$\sigma'_{xy}(0)$ divided by its maximum value, as a function of $\vphi$
for the $p$-wave gap and for the $f$-wave gap when
$\gamma/\Delta_{0}=0.05$. The maximum values are about $140\sigma_{0}$
and $420\sigma_{0}$ for the $p$-wave and $f$-wave case, respectively.
It is understandable that for a given $\vphi$
the magnitude of the DC Hall conductivity of the $f$-wave gap
is greater than that of the $p$-wave gap because
there are more quasiparticles in the $f$-wave case.
The strong dependence of $\sigma'_{xy}(0)$ on $\vphi$,
particularly between $\pi/2$ and $\pi$, makes the determination of $\vphi$ more
accurate.

Finally, it is necessary to address the chirality of
the $p$-wave and $f$-wave gap because it is interesting
to see if $\vphi$ changes this property.
The chirality\cite{volovik,furusaki} of a superconductor is
defined as
$
{\cal N} = \frac{1}{4\pi}\int{}d^{2}k\;
{\hat {\bf m}}\cdot\left(\frac{\partial {\hat {\bf m}}}{\partial k_{x}}
\times \frac{\partial {\hat {\bf m}}}{\partial k_{y}}\right)
$,
where ${\hat {\bf m}} = {\bf m}/|{\bf m}|$ with ${\bf m} =
(\mbox{Re}[\Delta_{\bk}],\mbox{Im}[\Delta_{\bk}],\xi_{\bk})$.
For the $p$- wave superconductor this becomes
\be
{\cal N}(\vphi)=-\int^{2\pi}_{0}\frac{d\theta}{2\pi}\frac{\sin(\vphi)}
{1+\cos(\vphi)\sin(2\theta)}\;.
\ee
Note that ${\cal N}(\vphi)=\pm1$ except for $\vphi=0$ and $\pi$,
for which the chirality is not uniquely defined because $\bf m$
vanishes at some points on the Fermi surface. The chirality
of the $f$-wave gap cannot be defined uniquely either because
$\bf m$ goes to zero at the nodal points on the Fermi surface.

The DOS of a superconducting state is defined as
$\frac{N(\om)}{N(0)}=
{\mbox{Re}}\left[
\left\la \frac{\om}{\sqrt{\om^{2}-|\Delta_{\bk}|^{2}}}\right\ra_{FS}
\right]$.
For the $p$-wave gap, we obtain
\be
\frac{N(\om)}{N(0)}=\frac{\pi}{2}{\mbox{Re}}\left[
\frac{\omega}{\sqrt{\om^{2}-2\Delta^{2}_{\vphi}}}\;
K\left(\frac{2\Delta^{2}_{0}\cos(\vphi)}{\om^{2}-2\Delta^{2}_{\vphi}}
\right)
\right]
\ee
where $\Delta_{\vphi}=\Delta_{0}\sin(\vphi/2)$.
Fig.~3 shows the DOS of the case with a $p$-wave gap.
It is interesting that the peak does not occur at $\omega=\Delta_{0}$
except for $\vphi=\pi/2$, for which the DOS is $s$-wave-like. In fact,
its location is $\omega/\Delta_{0}=\sqrt{1+|\cos(\vphi)|}$.
As mentioned early,
the DOS illustrates a tiny gap obtained in a different context\cite{miyake}.
When $\vphi\simeq5\pi/6$, the tiny (minimum) gap is about $0.3\Delta_{0}$
while the maximum gap (peak) about $1.37\Delta_{0}$.
Note that this value of $\vphi$ also
explains the Kerr angle measurement.
It is not possible to express the DOS of the case with $f$-wave gap for general $\vphi$
in an analytic form.
We plot the DOS in Fig.~4 for values of $\vphi=3\pi/4$, $4\pi/3$, and $\pi$.
The location of the peak is not $\om=\Delta_{0}$ either; for example,
the peak is at $\om/\Delta_{0}=(4/3)\sqrt{2/3}$ for $\vphi=\pi$.
As one can see, the DOS is more or less like the DOS of the $d$-wave gap.
When $\vphi=\pi/2$, the DOS is exactly $d$-wave-like.
However, for $\vphi=\pi$
the DOS is definitely not a linear function of $\om$ at low frequency
$(\om\ll\Delta_{0})$. This is due to the quadratic behavior of the corresponding $f$-wave
gap as a function of wave vector near the nodes. In addition,
because of this, the nodal approximation breaks down for $\vphi\approx \pi$.

{\it In conclusion},
we have proposed a novel type of superconducting order parameter symmetry, with
a relative phase $\vphi$ in the definition of the order parameter symmetry.
We then showed that the spontaneous Hall conductivity can be reduced by rotational symmetry breaking
as well as time reversal symmetry breaking. We have used the Kerr angle $\theta_{K}$
expression in terms of the Hall conductivity as in Ref.\cite{yakovenko}:
$\theta_{K}=(4\pi/\Omega)\mbox{Im}[\sigma_{xy}/(n^{3}-n)]$, where $n$ is the complex
index of refraction. It is apparent that the actual magnitude and phase of $n$ are important to determine
the angle. Another difficulty for a proper theoretical understanding is the issue of the detailed experimental setup discussed recently in Ref.\cite{lutchyn}. In fact, we think that the validity of the above expression for $\theta_{K}$ is still an open question for \sro. Our simple analysis is
an initial attempt to understand
the recent Kerr angle experimental results\cite{xia}. Because of these complications, a measurement of the low-frequency Hall conductivity is most desirable --- it will provide a definitive guideline for the theoretical modeling of \sro. Our $p$-wave model is compatible with the small gap due to the salient shape of the Fermi surface.

\begin{acknowledgments}

This work was supported in part by
NSERC, by ICORE (Alberta), by the Canadian Institute for Advanced Research
(CIfAR), and by the Robert Welch Foundation (grant no. E-1146). FM is grateful to the Aspen Center for
Physics, where some of this work was done. We thank Bob Teshima for technical help.

\end{acknowledgments}

\begin{figure}[tp]
\begin{center} \includegraphics[height=2.8in,width=2.8in]{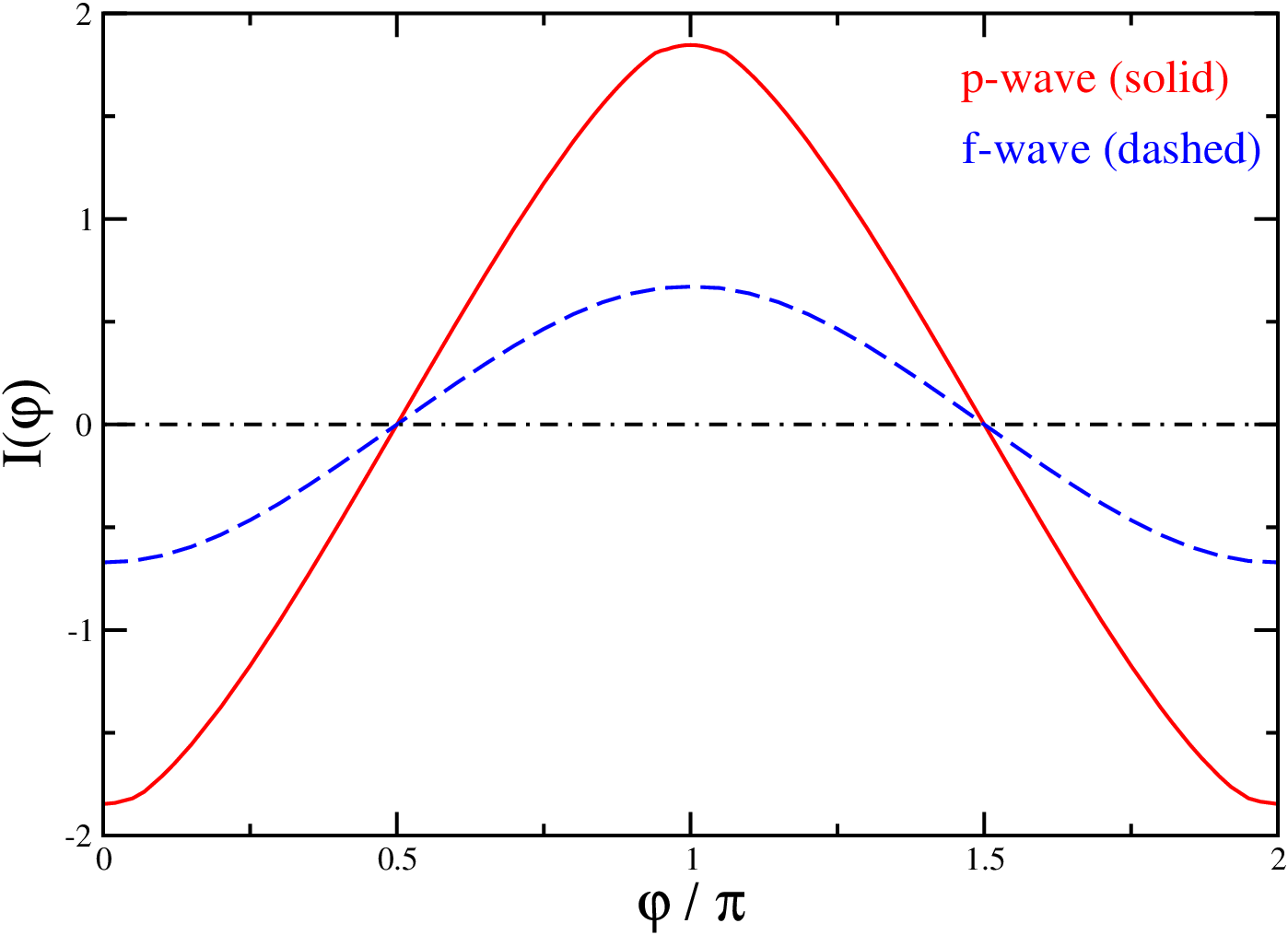}
\caption{(Color online)
$I(\vphi)$ for the $p$-wave gap (solid curve) and the $f$-wave gap (dashed curve).
}
\end{center}
\end{figure}

\begin{figure}[tp]
\begin{center} \includegraphics[height=2.8in,width=2.8in]{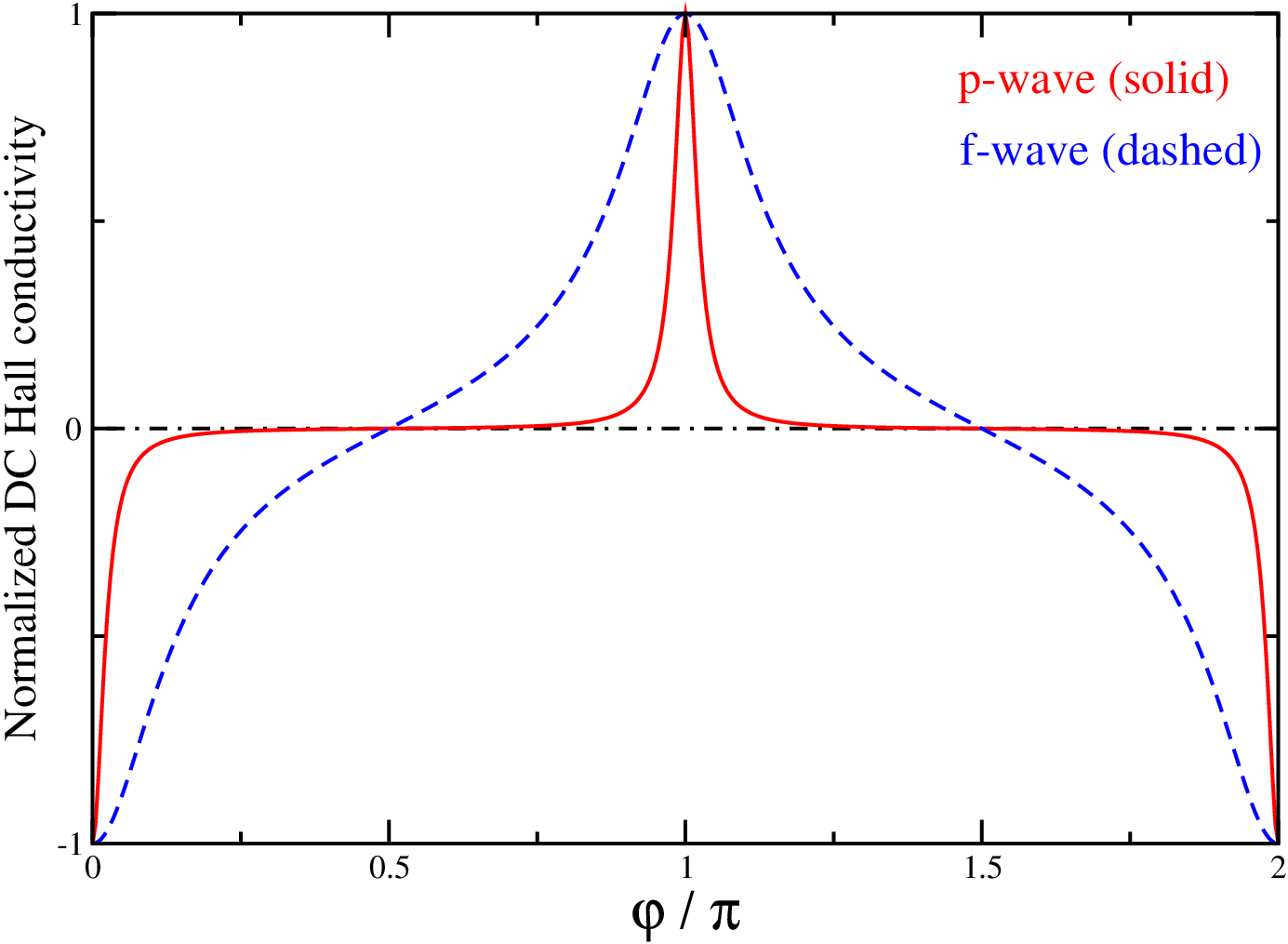}
\caption{(Color online) Normalized DC Hall conductivity as a function of
$\vphi$ at zero temperature
for $p$-wave and $f$-wave superconductor with $\gamma/\Delta_{0}=0.05$.
}
\end{center}
\end{figure}

\begin{figure}[tp]
\begin{center} \includegraphics[height=2.8in,width=2.8in]{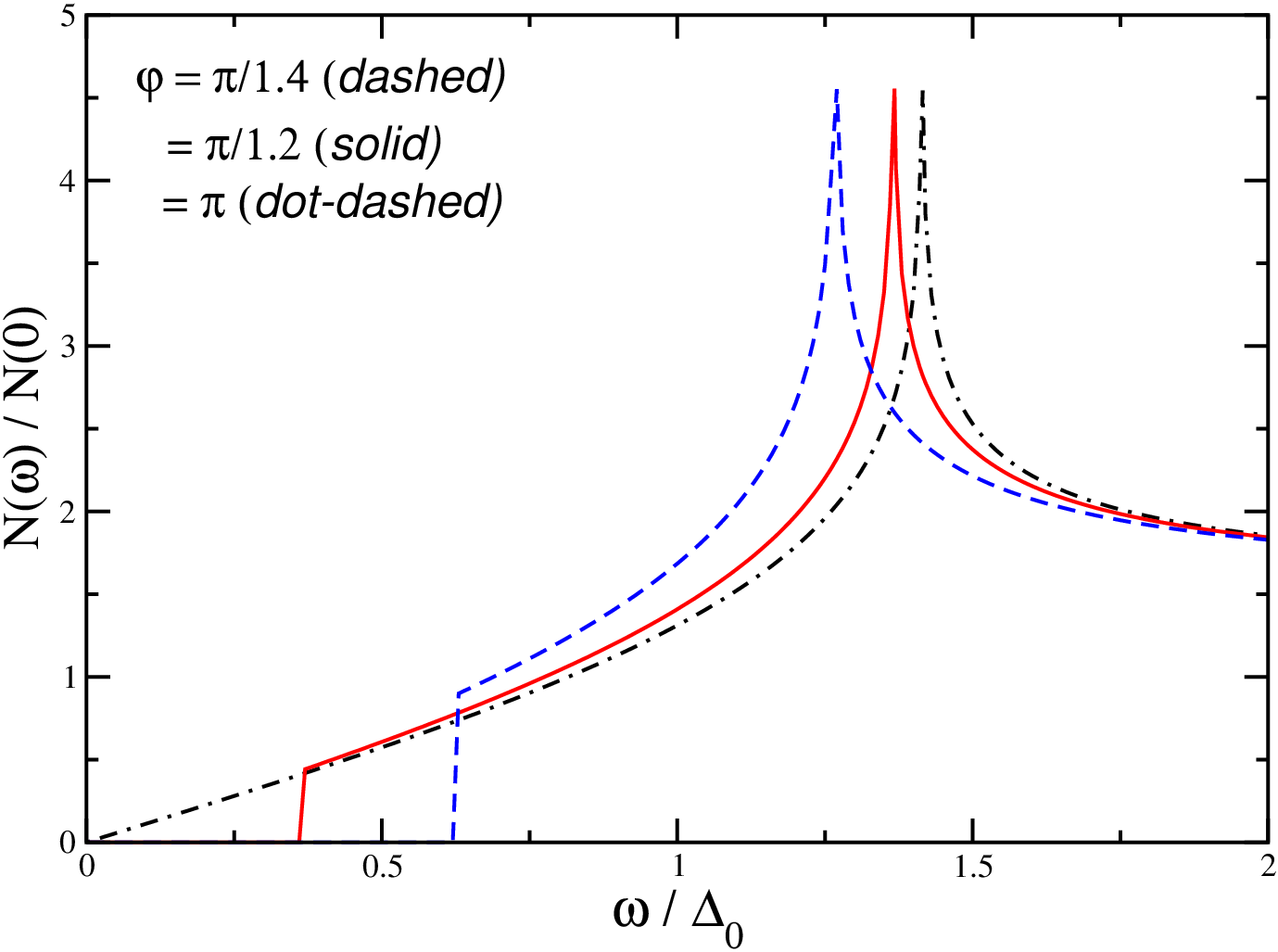}
\caption{The density of states of the $p$-wave gap as a function of energy $\om$.
}
\end{center}
\end{figure}

\begin{figure}[tp]
\begin{center} \includegraphics[height=2.8in,width=2.8in]{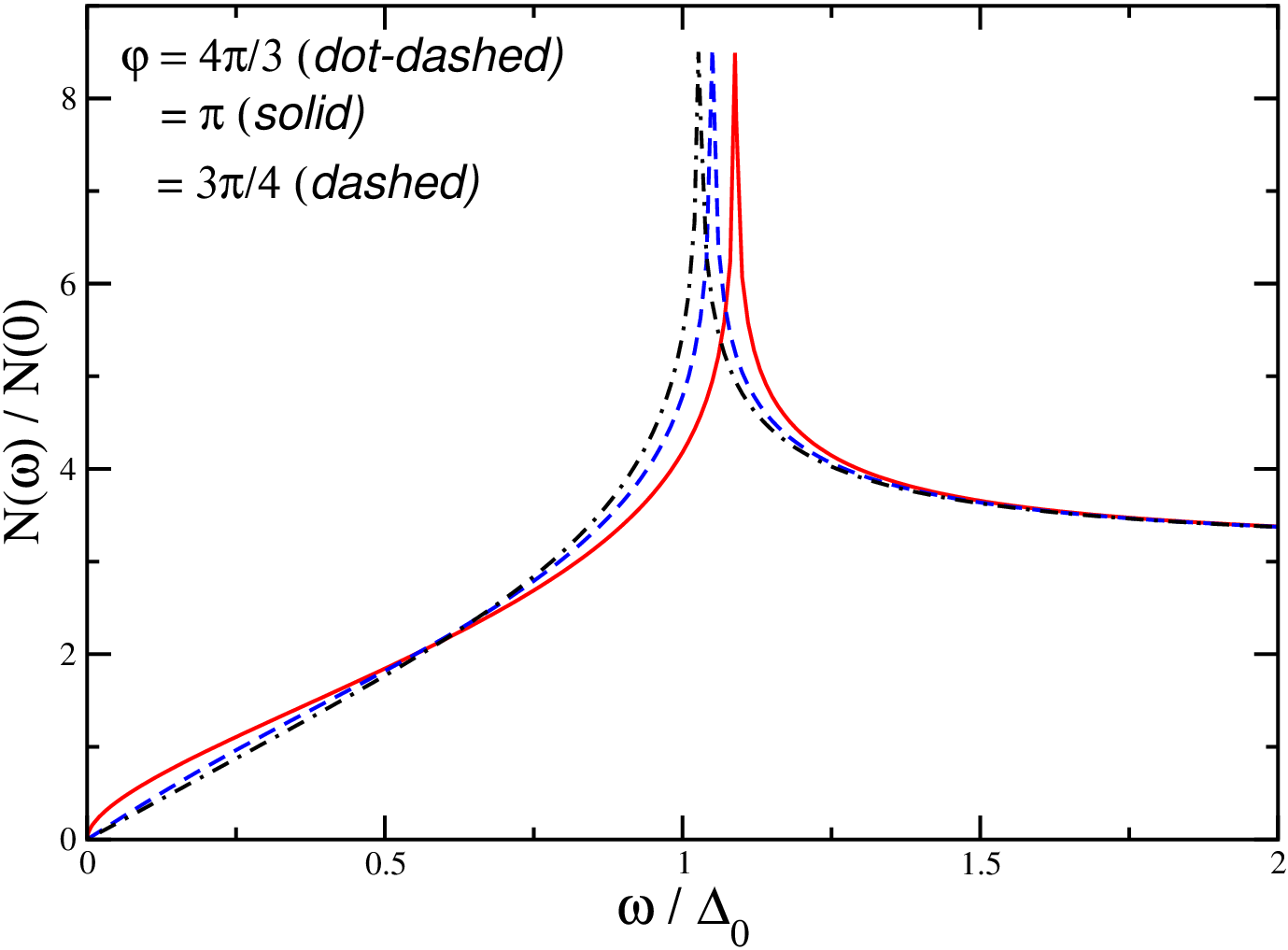}
\caption{The density of states of the $f$-wave gap as a function of energy $\om$.
}
\end{center}
\end{figure}

\bibliographystyle{prl}

\end{document}